\documentstyle[preprint,aps]{revtex}
\draft
\begin{document}
\title{Polaron band formation in the Holstein model}
\author{G. Wellein and H. Fehske } 
\address{Physikalisches Institut, Universit\"at Bayreuth,
D--95440 Bayreuth, Germany}
\date{Bayreuth, 24 February 1997}
\maketitle
\input{epsf}
\def\gsim{\hbox{$\lower1pt\hbox{$>$}\above-1pt\raise1pt\hbox{$\sim$}$}}
\def\lsim{\hbox{$\lower1pt\hbox{$<$}\above-1pt\raise1pt\hbox{$\sim$}$}}
\def\cH{{\cal{H}}}
\def\ep{\varepsilon_p}
\def\ho{\hbar\omega}
\begin{abstract}
We present numerical exact results for the polaronic band structure of the
Holstein molecular crystal model in one and two dimensions.  
The use of direct Lanczos diagonalization technique, preserving
the full dynamics and quantum nature of phonons, allows 
us to analyze in detail the renormalization of both 
quasiparticle bandwidth and dispersion by the 
electron--phonon interaction. 
For the two--dimensional case some of our exact data are compared 
with the results obtained in the framework of a recently developed 
finite cluster strong--coupling perturbation theory.    
\end{abstract}
\thispagestyle{empty}
\noindent
The very fundamental problem of a single conduction electron  
coupled to bosonic degrees of freedom is still not 
completely understood. In the case where the bosons are 
lattice phonons that is what is usually called the polaron 
problem and numerous analytical techniques
were tried out to tackle this complicated many--body 
problem in terms of the simplest electron--phonon (EP) 
Hamiltonians~\cite{Mah90}. Depending on the relative 
importance of the long--range or short--range
electron--lattice interactions simplified models of the Fr\"ohlich~\cite{Fr54}
or Holstein~\cite{Ho59a} type, respectively, have been widely used to analyze 
polaronic effects in solids with displaceable atoms. 
The crucial point is that the usual phase transition concept 
fails to describe the crossover from an only weakly dressed electron 
to the strongly renormalized less--mobile polaronic quasiparticle 
with increasing EP coupling strength. 
Furthermore, as yet none of the various  analytical treatments,   
based on variational approaches~\cite{Em73} or on  
weak--coupling~\cite{Mi58} and strong--coupling 
adiabatic~\cite{Ho59a} and non-adiabatic~\cite{LF62,Ma95} 
perturbation expansions, are suitable for the investigation 
of the physically most interesting transition region, 
where the highly non--linear ``self--trapping'' process 
of the charge carrier takes place.  
That is because precisely in this situation 
the characteristic electronic and phononic energy scales 
are not well separated and non--adiabatic effects 
become increasingly important, implying a
breakdown of the standard Migdal approximation~\cite{AM94b}. 
In principle, quasi approximation--free numerical methods like
(quantum) Monte Carlo simulations~\cite{RL83} 
and exact diagonalizations (ED)~\cite{Fe96} 
can close the gap between the weak and strong EP coupling limits
and therefore, at the moment, provide the only reliable 
tool for studying polarons close to the crossover regime. 
Along this line, previous ED work has concentrated on 
spectral properties of the one--dimensional (1D) Holstein  
polaron. In practice, however, memory limitations have restricted
the calculations to either small values of the EP 
coupling constant or to very small clusters~\cite{RT92,Ma93,AKR94},
thus, e.g., making a more detailed discussion of the polaronic band 
dispersion impossible. First ED studies performed on the 1D ten--site
Holstein model indicate that the small polaron dispersion may differ
significantly from a rescaled cosine tight--binding band especially
in the crossover regime~\cite{WRF96}; a result which has been 
corroborated more recently by a finite cluster (FC) strong--coupling 
perturbation theory~(SCPT)~\cite{St96}.       

Encouraged by these findings, it is the aim of the present paper
to perform a systematic Lanczos study of the 1D and 2D Holstein model
on large enough lattices, in order to discuss 
the evolution of the polaronic quasiparticle band structure 
in dependence on the phonon frequency and EP interaction strength.

The Holstein Hamiltonian reads:
\begin{equation}
\cH=-t\sum_{\langle ij\rangle} ( c_i^{\dagger} c_j^{} 
+ c_j^{\dagger} c_i^{}) -\sqrt{\ep\ho}\sum_i
( b_i^{\dagger}  + b_i^{})  c_i^{\dagger} c_i^{}
+\ho\sum_i  ( b_i^{\dagger} b_i^{}+ 
\mbox{\small $\frac{1}{2}$})\,,
\end{equation}
where $ c_i^{[\dagger]}$ and $ b_i^{[\dagger]}$ are the annihilation 
[creation] operators of a (spinless) fermion and a boson (phonon) 
at Wannier site $i$, respectively. In~(1), the following idealizations
of real electron--phonon systems are made: (i) the electron 
transfer $t$ is restricted to {\it nearest--neighbour} (NN) pairs
$\langle ij\rangle$; 
(ii) the charge carrier is {\it locally}
coupled to a {\it dispersionsless} optical phonon mode 
($\ep$ denotes the EP coupling constant and $\hbar\omega$ is the 
bare phonon frequency); 
(iii) the phonons are treated within an {\it harmonic} approximation.
Recall that the physics of the Holstein model 
is governed by {\it two} dimensionless
parameters $\lambda=\ep/2Dt$ and $\alpha =\sqrt{\ep/\hbar\omega}$; i.e., 
irrespective of the adiabaticity ratio $\gamma=t/\hbar\omega$,  
small polarons are formed provided that $\lambda > 1$ {\it and} 
$\alpha >1$.   

In the numerical analysis of the Holstein model, we employ the 
standard Lanczos algorithm in combination with a  
well--controlled  truncation of the phononic Hilbert space. 
According to the truncation procedure described in~\cite{WRF96}, we
have to check for the convergence of both the ground--state energy $E_0(M)$
and the phonon distribution function, which has to be
 determined self--consistently in the ground state $|{\mit \Psi}_0(M)\rangle$. 
Increasing the maximum number of phonons retained $(M)$, 
convergence is assumed to be achieved   
if the relative error $|E_0(M+1)-E_0(M)|/|E_0(M)|$ 
becomes less than $10^{-7}$. 

In order to discuss the single--polaron polaron band structure, 
we make use of translational invariance and compute the 
wave--vector resolved spectral density function  
\begin{equation}
  A_{\vec{K}}(E) = \sum_n |\langle {\mit\Psi}_{n,\vec{K}}^{(1)} 
\,|\,c_{\vec{K}}^{\dagger}\,|\,0\rangle|^2\,\delta ( E-E_n^{(1)})\,,
\end{equation}
using a polynomial moment expansion together with the 
maximum entropy method~\cite{Siea96}. 
Then the so--called ``coherent'' band dispersion, $E_{\vec{K}}$, is obtained 
from the first peak of $A_{\vec{K}}(E)$ having finite spectral weight 
\begin{equation}
Z_{\vec{K}}= |\langle {\mit\Psi}_{0,\vec{K}}^{(1)} 
\,|\,c_{\vec{K}}^{\dagger}\,|\,0\rangle|^2\,,
\end{equation}
where $|{\mit \Psi}_{0,K}^{(1)}\rangle$ denotes the single--electron 
state being lowest in energy in a certain $\vec{K}$--sector. 

In the first place, we focus on the 1D model and start with 
the study of the {\it weak--coupling case} ($\lambda < 1$). 
Figure 1 displays the band dispersion calculated
at $\ep=0.1$ for various phonon frequencies ranging 
from the adiabatic $(\gamma =10)$ to the non--adiabatic
$(\gamma =0.125)$ regime (in the following all energies
are measured in units of $t$). For low and intermediate 
phonon frequencies, the energy to excite one phonon lies inside
the bare tight--binding band, $E^{0]}_{\vec{K}}=-2t\sum_{i=1}^D
\cos K_i$, of a D--dimensional hypercubic lattice with unit lattice spacing.   
Thus at arbitrarily small $\ep$, predominantly ``phononic'' states enter 
the low--energy spectrum in those $\vec{K}$--sectors for which 
$E^{[0]}_{\vec{K}}$ being separated from the ground state 
by an energy $E\stackrel{>}{\sim} \hbar\omega$. 
As a result the dispersion curves reflect the well--known
peculiarities of the absorption spectra of an electronic or excitonic system 
weakly interacting with dispersionsless optical phonons~\cite{ES63,Sh87},
i.e., a nearly unaffected cosine dispersion near the band center ($\vec{K}=0$)
and a practically flat region at larger momenta. Note that even in cases where
the phonon frequency is comparable to the bare electronic 
bandwidth a remarkable  ``flattening'' of the band structure 
appears in the vicinity of the zone boundary 
(e.g., at $\hbar\omega =4$ we found 
$E_{\pi}(\ep=0.1)/E_{\pi}(0)\sim 0.92$). 
In the extreme non--adiabatic regime ($\gamma \ll 1$) the 
phonon distribution function is sharply peaked at the 
zero--phonon state and, because the interaction is weak
and the phonons can follow the electron instantaneously,  
we obtain a negligible renormalization of the band structure. 
To substantiate this interpretation we have 
calculated the  $\vec{K}$--dependent mean phonon number,
\begin{equation}
N^{ph}_{\vec{K}}= \sum_i \langle {\mit\Psi}_{0,\vec{K}}^{(1)}\,|\,
b_i^{\dagger} b_i^{} \,|\, {\mit\Psi}^{(1)}_{0,\vec{K}} \rangle\,,
\end{equation}
in the lowest band states $|{\mit\Psi}^{(1)}_{0,\vec{K}}\rangle$.
The results are presented in the inset of Fig.~1 for three characteristic 
momenta $\vec{K}=0,\;0.4\pi$, and $\pi$. Evidently the 
states keeping small momenta are basically zero--phonon states.
For larger momentum, we observe a gradual crossover from  one--phonon 
to zero--phonon states upon increasing the phonon frequency, 
whereby  the admixture of ``electronic character'' increases.

Of course it is of special interest to understand  
the evolution of the band structure by increasing the  
EP coupling strength.  Thus it is fruitful to compare $E_{\vec{K}}$   
for the weak and strong--coupling cases at fixed phonon frequency.
This has been done in Fig.~2 for various system sizes $N=10, \dots, 20$,
using EP couplings $\ep=0.5$ (a) and $\ep =4.0$ (b) together with an 
intermediate adiabaticity parameter $\gamma= 1.25$. 
First of all one notices that the finite--size effects 
are very small, i.e., although the data points belong to different
lattice sizes we obtain a remarkably smooth behaviour of $E_{\vec{K}}$. 
As might be expected, for $\ep=0.5$ the coherent bandwidth,  
${\mit \Delta}E=\sup_{\vec{K}} E_{\vec{K}}-\inf_{\vec{K}} E_{\vec{K}}$,
is approximately given by the phonon energy 
(${\mit \Delta}E=0.782\sim\hbar\omega=0.8$) 
and the dispersion again shows the coexistence
of two different types of band states at small and large momenta. 
By contrast, if the EP interaction is enhanced, small polaron 
formation takes place and we observe all signs of the famous polaronic 
band collapse. However, note that even in 
the relatively strong EP coupling regime
displayed Fig.~2~(b) the standard Lang--Firsov 
formula, ${\mit \Delta}E_{LF}=4Dt\exp[-\ep/\hbar\omega]$ 
(obtained by performing the 
Lang--Firsov canonical transformation~\cite{LF62} and taking 
the expectation value of the kinetic energy over the transformed 
phonon vacuum), does not give a satisfactory estimate
of the bandwidth. So we found ${\mit \Delta}E_{LF}=0.0269$ which has to be 
contrasted  with the exact result ${\mit \Delta}E=0.0579$. 
Besides the band narrowing effect, there are several 
other features worth mentioning for polaronic band states 
in the crossover region. 
Most notably, throughout the whole Brillouin zone
the band structure differs significantly from that of a 
rescaled tight--binding (cosine) band containing only NN hopping. 
As can be seen from a least--squares fit to an effective band 
dispersion $\bar{E}_{\vec{K}}=\sum_{l=0}^3 a_l \cos l K$, the residual 
polaron--phonon interaction generates {\it longer--range} 
hopping terms~\cite{WRF96,Fi75}. Concomitantly, the 
mass enhancement due to the EP interaction is weakened at the band minimum. 
It is important to realize that these effects are most pronounced 
at {\it intermediate} EP couplings and phonon frequencies. In this parameter 
region even  higher--order SCPT, with its
internal states containing some excited phonons,  seems to be not
tractable because the convergence of the series 
expansion is  poor~\cite{St96}. 
The deviation from the cosine dispersion at different $\ep$ is
depicted in the inset of Fig.~2~(a), where we have
used the eight and ten site lattices in order to reach larger
EP coupling strengths. Here $E_{\vec{K}}$ is scaled with respect to
the coherent bandwidth which strongly depends on $\ep$ and $\hbar\omega$,
for example, for $N=8$ and $\hbar\omega=0.8$ 
we found ${\mit \Delta}E(\ep)=0.1957$,
0.0559, and 0.0143 for $\ep=3.0$, 4.0, and 5.0, respectively.
It is obvious and doesn't need discussion that the 
dispersion is barely changed from a rescaled tight--binding band   
in the very extreme small polaron limit ($\lambda \gg 1$).
To visualize in more detail the different nature of 
large and small polaron band states, we have evaluated numerically  the 
quasiparticle residue $Z_{\vec{K}}$ and the mean phonon number 
$N_{\vec{K}}^{(ph)}$
at all  $\vec{K}$--points. Obviously, the $\vec{K}$--dependent 
renormalization factor can be taken as a measure of the ``electronic
contribution'' to the polaronic quasiparticle. The results are
shown in Fig.~2~(c). For weak EP coupling, one recognizes immediately the
electronic and phononic character of the states at small and large
momentum, respectively. The crossover between these two types of states 
is accompanied by a change in the mean phonon number by approximately one. 
With increasing $\ep$ a strong mixing of electron and phonon degrees of
freedom takes place, whereby, forming a small polaron, 
both quantum objects completely loose their own identity. 
As expected, this  leads to a significant suppression 
of the quasiparticle residue for all~$\vec{K}$. 
At large $\ep$, the polaronic state is 
characterized by strong on--site electron--phonon 
correlations making the quasiparticle susceptible to self--trapping. 
Thus the coherent movement of the carrier is greatly hindered
but itinerant band states do still exist. 
The small polaron is heavy because it has to drag with it   
a large number of phonons in the so--called ``phonon cloud''
(cf. Fig.~2~(c) inset). Since the high--energy band edge states  
are more vulnerable to decay from EP interaction
the particle loses its spectral weight more rapidly at
$\vec{K}=\pi$~\cite{KMKF96}. 

Next, we consider the evolution of the band
structure for the two--dimensional case.  Figure~3 presents the results
obtained for $E_{\vec{K}}$ and $Z_{\vec{K}}$ 
along the highly symmetric directions 
of the Brillouin zone in the weak (a) and intermediate 
(b) coupling regime. As in the (weak--coupling)
1D case, we found a flattening of the band structure in 
regions where $E_{\vec{K}}^{[0]}$ becomes comparable to the phonon energy.  
The sharp drop in the quasiparticle weight obtained with increasing
$|\vec{K}|$ at small $\hbar\omega$ is weakened in the non--adiabatic regime.
That is, the transition region $(0.25\stackrel{<}{\sim}Z_{\vec{K}}
\stackrel{<}{\sim} 0.75 )$ is enlarged. 
In order to make a comparison with recent predictions of both 
second--order Rayleigh--Schr\"odinger SCPT and 
the re-summed FC--SCPT developed by Stephan~\cite{St96},
we have plotted  $E_{\vec{K}}$ and $Z_{\vec{K}}$ for
$\ep=5.0$ and $\hbar\omega=1.5$
in the lower part of Fig.~3 . Interestingly enough,   
the flattening of the dispersion at the edges of the Brillouin zone
survives to surprisingly strong EP couplings (note that
$\hbar\omega \stackrel{>}{\sim} 2 {\mit \Delta} E$). 
The wave--vector renormalization factor $Z_{\vec{K}}$ clearly indicates 
the polaronic nature of the charge carrier. That means, calculating the
strong--coupling wave--function we have to include multiphonon states. 
Once again the Lang--Firsov formula (first--order SCPT) 
underestimates the exact bandwidth by more than a factor 
of two; ${\mit \Delta}E_{LF}=0.2854$. Even if we include second--order
corrections the  SCPT band structure significantly deviates from 
the exact result. Obviously, the agreement is poorest 
at the $\Gamma$--point ($\vec{K}=0)$. Thus we are lead to the conclusion
that the standard strong--coupling approximation fails to accurately
predict the ground state energy and the polaronic bandwidth 
in the moderate strong--coupling regime. 
By contrast, the FC--SCPT approach~\cite{St96},
which in our case makes use of information extracted from a 
sequence of finite--cluster diagonalizations up to five sites, provides 
a correct description of the coherent band structure.
The extremely good success of the FC--SCPT at intermediate
EP couplings and phonon frequencies may be attributed to the inclusion of 
longer--range correlations, which are especially important in the
crossover region from nearly free to small polarons.

To summarize, we have performed a comprehensive study of 
the 1D and 2D Holstein model using exact diagonalizations.  
Concerning the polaronic band structure of the Holstein model,
our main results are the following. 

(i) In the weak--coupling case and for phonon frequencies less 
than the bare electronic bandwidth, we recover the expected
flattening of the band dispersion at large momentum. Here our 
numerical results for the wave--vector resolved quasiparticle weight factor 
and the mean phonon number clearly demonstrate the electronic and phononic 
character of the band states near the band center and band edges, 
respectively. 

(ii) Increasing the electron--phonon coupling the flattening 
survives to a large extent the crossover from large to small 
polarons. The small polaron state is characterized by less 
spectral weight and a large average number of phonons.  

(iii) At intermediate EP coupling strengths and phonon frequencies
the effective polaronic band dispersion deviates substantially from a simple 
tight--binding cosine band due to further than nearest--neighbour ranged 
hopping processes generated by the residual polaron--phonon interaction,
implying the importance of multiphonon states. In the intermediate
coupling regime, the polaronic mass enhancement can be
one order of magnitude smaller than predicted by standard 
strong--coupling perturbation theory.   

(iv) In the extreme strong--coupling limit $(\ep\gg \hbar\omega,\,t)$, 
we have a nearly dispersionsless small polaron band with almost vanishing 
spectral weight. That means, here the exponential band narrowing 
completely dominates the physics of the Holstein model and, in real solids, 
coherent polaron motion will be immediately destroyed by disorder effects.  

The authors are greatly indebted to W.~Stephan for
putting his 2D FC--SCPT data at our disposal. 
H.~F. is especially grateful to D.~Ihle and J.~Loos for many 
useful discussions. This work was performed  under the auspices of Deutsche 
Forschungsgemeinschaft, SFB 279, Bayreuth.
Special thanks go to the LRZ M\"unchen for the generous granting of 
their computer facilities.
\narrowtext\def\baselinestretch{0.95}
\bibliography{ref}
\bibliographystyle{phys}
\figure{FIG. 1. 
Polaronic band dispersion $E_{\vec{K}}$ for the 1D Holstein model. Data points
are exact results obtained for a finite lattice with $N=20$ sites 
and at most $M=10$ phonons using periodic boundary conditions. 
Here and in the following the energy scales of the ordinates  
were shifted by $N\hbar\omega/2$.  The inset shows the mean phonon number 
$N^{ph}_{\vec{K}}$ in band states with wave--vector $\vec{K}$.} 
\figure{FIG. 2.
Polaronic band structure of the 1D Holstein model in the
weak (a) and strong (b) coupling regimes, where the phonon
frequency is $\hbar\omega=0.8$. For both cases, 
the quasiparticle residue and mean phonon number are given
in (c).}
\figure{FIG. 3. Band dispersion $E_{\vec{K}}$ (filled symbols) 
and quasiparticle weight factor $Z_{\vec{K}}$ (open symbols) 
for the 2D Holstein model with $\ep=0.5$, $\hbar\omega =0.8$ (circles),    
4.0 (squares) and  $\ep=5.0$, $\hbar\omega = 1.5$ (diamonds).
ED results are obtained using finite square lattices with $N=16$ and
18 sites (the lines are only a guide for the eye). 
In the weak--coupling case (a) the 
chain dashed curve gives the free electron dispersion $E_{\vec{K}}^{[0]}$; 
for the strong--coupling case (b) exact 
data (filled diamonds) are compared with the predictions of  
standard second--order SCPT (dashed line) and 
FC--SCPT (solid line)~\protect\cite{St96}.}  
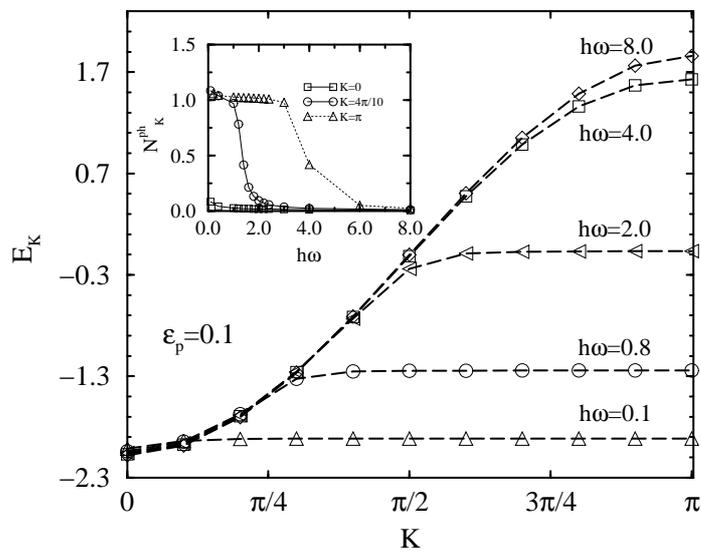
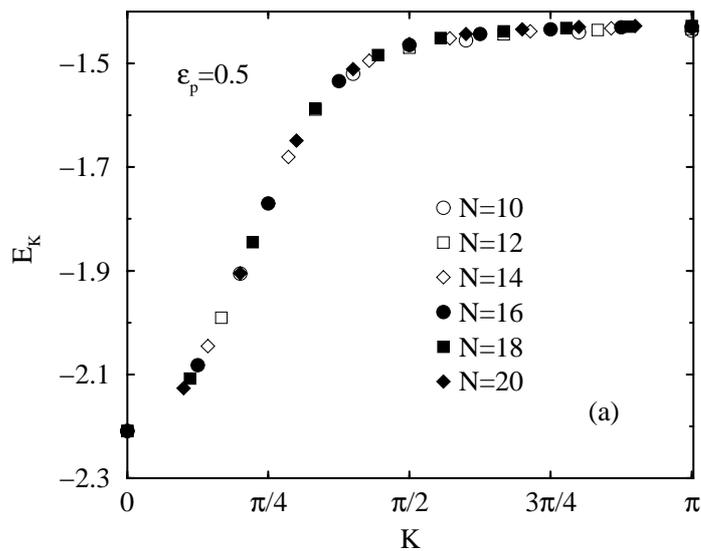
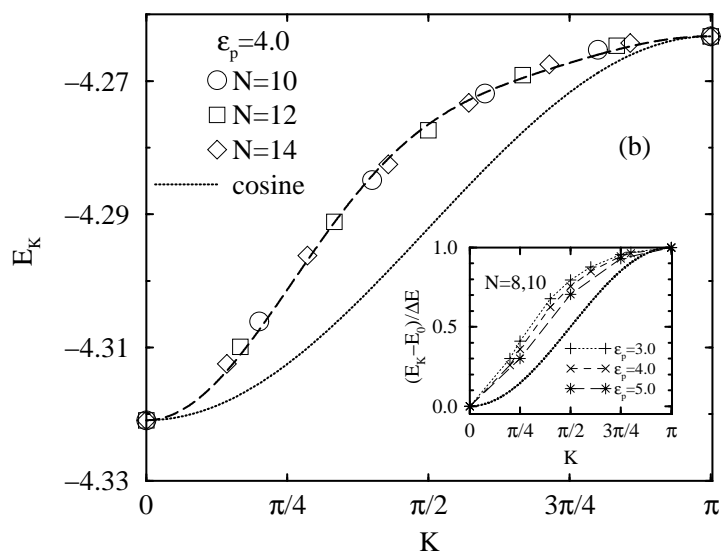
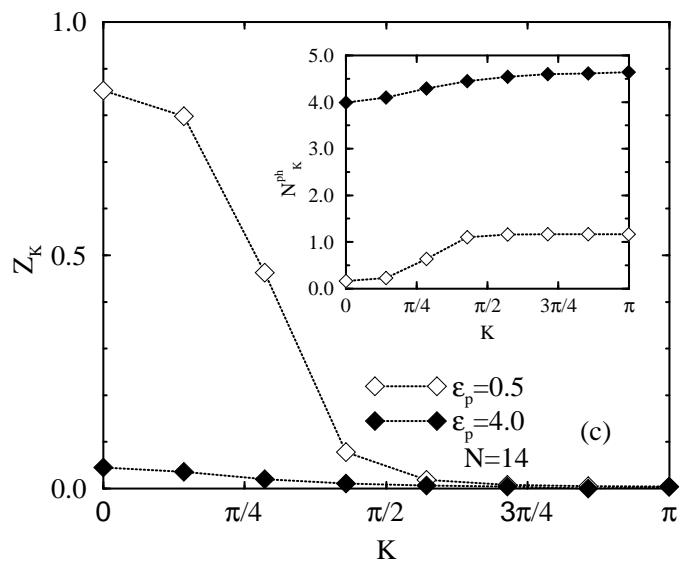
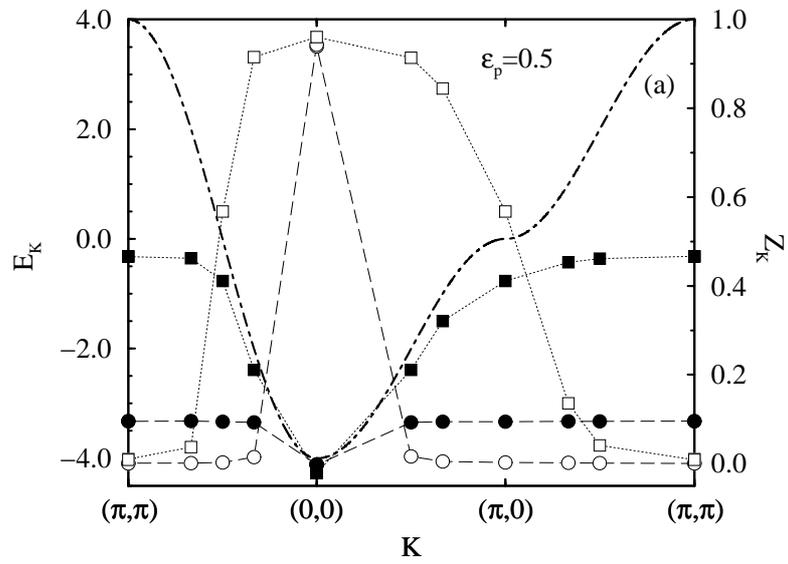
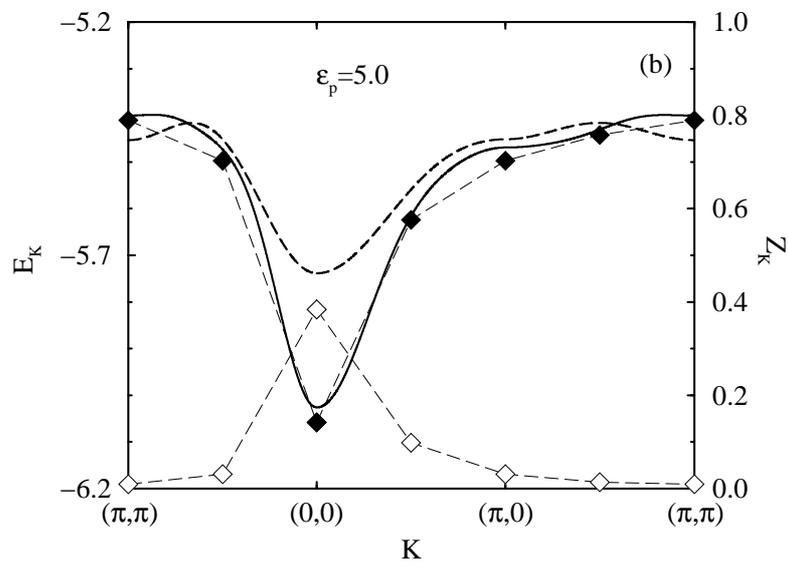
\end{document}